\documentstyle[aps,psfig]{revtex}
\newcommand{\be}{\begin{equation}}
\newcommand{\ee}{\end{equation}}
\newcommand{\ben}{\begin{eqnarray}}
\newcommand{\een}{\end{eqnarray}}
\newcommand{\ov}{\overline}
\begin{document}

\title{Complete Factorization of Equations of Motion in
Wess-Zumino Theory}
\author{D. Bazeia,$^a$ J. Menezes,$^a$ and M. M. Santos$^b$}
\address{$^a$Departamento de F\'\i sica, Universidade Federal
da Para\'\i ba, Caixa Postal 5008, 58051-970 Jo\~ao Pessoa PB,
Brazil\\$^b$Departamento de Matem\'atica, Universidade Estadual de
Campinas, Caixa Postal 6065, 13081-970 Campinas SP, Brazil}
%\date{\today}

\maketitle

\begin{abstract}
We prove that the equations of motion describing domain walls in a
Wess-Zumino theory involving only one chiral matter multiplet can
be factorized into first order Bogomol'nyi equations, so that all
the topological defects are of the Bogomol'nyi-Prasad-Sommerfield
type.
\end{abstract}

\bigskip


\bigskip

Domain walls are defect structures that appear in diverse branches
of physics. They usually live in three spatial dimensions as
bidimensional objects that arise in systems described by
potentials that contain at least two isolated degenerate minima.
They involve energy scales as different as the ones that appear in
Condensed Matter \cite{cm} and in Cosmology \cite{cos}.

A lot of attention has been drawn recently to domain walls in
field theories, in models that have been investigated under
several distinct motivations
\cite{fmv90,ato91,cqr91,sg,wit,95,97,dsh97,sve97,gto99,bbr00,bve00,nol00}.
A very specific motivation concerns the presence of domain walls
arising in between non zero vacuum expectation values of scalar
fields in supergravity \cite{cqr91,sg}. Another line deals with
the formation of defects inside domain walls \cite{wit,95,97}. A
great deal of attention has also been drawn to $SU(N)$
supersymmetric gluodynamics, where nonperturbative effects give
rise to gluino condensates that may form according to a set of $N$
isolated degenerate chirally asymmetric vacua, from where domain
walls spring interpolating between pairs of vacua
\cite{dsh97,sve97}. The interest in domain walls in general widens
because of the interplay between Field Theory and the low energy
world volume dynamics of branes in String Theory
\cite{sb,sb1,sb2,sbf}.

In the present work we examine the bosonic portion of a
Wess-Zumino theory that involves only one chiral matter multiplet.
We write the Lagrangian density in the standard form
\ben
{\cal L}=\frac12\,\partial_{\alpha}{\ov\varphi}\,\partial^{\alpha}\varphi-
\frac12|W'(\varphi)|^2.\nonumber
\een
We assume that the superpotential $W({\varphi})$ is a holomorphic function
in the complex variable $\varphi$, and $W'(\varphi)$ stands for the complex
derivative of $W$ with respect to $\varphi$.
The specific form of the potential shows that its minima
are critical values of the superpotential.

We search for defect structures, for static solutions of the
equation of motion. We suppose the static solutions depend only on
$x$, so the equation of motion becomes
\be
\label{soceq}
\frac{d^2\varphi}{dx^2}=W'(\varphi)\,{\ov{W''(\varphi)}}.
\ee
We examine the energy of the static configurations to show that the
second order ordinary differential equation is
equivalent to a family of first order ordinary differential equations
of Bogomol'nyi type, namely
\be
\label{foceq}
\frac{d\varphi}{dx}={\ov{W'(\varphi)}}\,e^{-i\xi},
\ee
where $\xi$ is the family parameter ranging in the set of real
numbers, constant with respect to $x$. Thus, we say
that the equation of motion (\ref{soceq}) completely factorizes
into a family of first order equations.
This is our main result. It is valid under the boundary conditions
$\lim_{x\to-\infty}\varphi(x)={v^k}$,
$\lim_{x\to-\infty}(d\varphi/dx)=0$, as required by the
topological solutions, where $v^k$ is some critical point of the
superpotential $W$. A direct consequence of this result is that in such
models all the topological solutions are of the BPS type, that is, these
models do not support non-BPS states.

Next, we proceed to prove our main result, that is, the equivalence
of the equation of motion (\ref{soceq}) and the family of first
order equations (\ref{foceq}). It is obvious that (\ref{foceq})
implies (\ref{soceq}), for any $\xi$, as one can see just by differentiating
equation (\ref{foceq}) with respect to $x$. To prove the reverse, that is,
to prove that any solution of (\ref{soceq}) under the aforementioned boundary
conditions is a solution of (\ref{foceq}) for some $\xi$, we introduce the
ratio
\ben
R(\varphi)=\frac{1}{\ov{W'({\varphi})}}\,\frac{d\varphi}{dx}\, .\nonumber
\een
Differentiating $R(\varphi)$ with respect to $x$ and using the
equation of motion (\ref{soceq}), we obtain
\ben
\frac{dR(\varphi)}{dx}&=&\left(\frac1{\ov{W'(\varphi)}}\right)^2\Biggl[
\,{\ov{W'(\varphi)}}\,\frac{d^2\varphi}{dx^2}-
\frac{d\varphi}{dx}\,{\ov{W''(\varphi)}}\,
\frac{d{\ov\varphi}}{dx}\Biggr]\nonumber
\\
&=&\left(\frac1{{\ov{W'(\varphi)}}}\right)^2\Biggl[
\Bigl|W'(\varphi)\Bigr|^2-\Bigl|\frac{d{\varphi}}{dx}
\Bigr|^2\Biggr]\,{\ov{W''(\varphi)}}\, .\nonumber
\een
We notice that
\ben
\label{eq1}
\frac{d}{dx}|W'(\varphi)|^2
={\ov{W'(\varphi)}}W''(\varphi)\frac{d\varphi}{dx}
+W'(\varphi){\ov{W''(\varphi)}} \frac{d{\ov\varphi}}{dx}
=\frac{d}{dx}\biggl|\frac{d\varphi}{dx}\biggr|^2.\nonumber
\een
Besides, due to the boundary conditions we also have
$\lim_{x\to-\infty}W'(\varphi)=0$. Thus
\be
|W'(\varphi)|^2-\biggl|\frac{d\varphi}{dx}\biggr|^2=0,
\label{eq2}
\ee
and so, $R(\varphi)$ does not depend on $x$. It remains only to show
that $R(\varphi)$ has modulus one. This is a consequence of
(\ref{eq2}) again. Indeed,
\ben
|R(\varphi)|^2=R(\varphi)\ov{R(\varphi)}
=\frac{1}{|W'(\varphi)|^2}\biggl|\frac{d\varphi}{dx}\biggr|^2=1.\nonumber
\een

We now present our last remarks. First, our result is valid in the case
of Wess-Zumino theory involving only one chiral matter multiplet,
so it does not work for models with more than one multiplet. Second,
in systems of scalar fields one may occasionally run into solutions
that engender no topological feature. They are named
nontopological solutions, and have been found for instance in
Ref.~{\cite{mrw}}. We first notice that in systems of real scalar
fields described by some superpotential, the nontopological
solutions cannot appear as solutions of the first order
Bogomol'nyi equations, because they should have zero energy, and
this is the energy of the vacuum states. We add this to our former
result to obtain another result: in the bosonic sector of a Wess-Zumino
theory involving only one chiral matter multiplet there is no
room for non topological solutions. Third, if we
deal with a single real scalar field $\phi$, we have to consider
potentials in the form $V(\phi)=(1/2)\,W^2_{\phi}$. In this case
the equation of motion for $\phi=\phi(x)$ becomes
$d^2\phi/dx^2=W_{\phi}\,W_{\phi\phi}$. It also factorizes into the
two first order Bogomol'nyi equations $d\phi/dx=\pm \,W_{\phi}$,
for solutions that obey $\lim_{x\to-\infty}\phi(x)\to{v^k}$ and
$\lim_{x\to-\infty}(d\phi/dx)=0$, where ${v^k}$ is a minimum of
$V(\phi)$, a critical point of $W(\phi)$.

In summary, the new idea of factorizing equations of motion into
first order Bogomol'nyi equations is of direct interest to
supersymmetry. It can be used in diverse applications, in
particular to investigate the presence of BPS walls interpolating
between distinct pairs of vacua, and exact integrability. The
complete factorization of equations of motion in systems described
by $W(\varphi)$ holomorphic ensures BPS feature to domain walls
that spring in such systems. Because of the associated BPS
character, these domain walls partially preserve the
supersymmetry.

We thank F.A. Brito, J.R. Morris, J.R.S. Nascimento, and R.F.
Ribeiro for discussions, and CAPES, CNPq and PRONEX for partial support.

\end{document}